\documentclass[a4paper]{spie}
\usepackage{graphicx}
\usepackage{amsfonts,amssymb,amsbsy,amsmath}

\title{Position-dependent photon operators in the quantization of the electromagnetic field in dielectrics at local thermal equilibrium} 
\author{Mikko Partanen, Teppo H\"ayrynen, Jani Oksanen, and Jukka Tulkki\skiplinehalf
Department of Biomedical Engineering and Computational Science,\\Aalto University, P.O. Box 12200, 00076 Aalto, Finland}

\begin{document} 

\maketitle 

\begin{abstract}
It has very recently been suggested that asymmetric coupling of electromagnetic
fields to thermal reservoirs under nonequilibrium conditions can produce unexpected
oscillatory behavior in the local photon statistics in layered structures.
Better understanding of the predicted phenomena could enable useful applications
related to thermometry, noise filtering, and enhancing optical interactions.
In this work we briefly review the field quantization and study the local steady
state temperature distributions in optical cavities formed of lossless and lossy
media to show that also local field temperatures exhibit oscillations that depend
on position as well as the photon energy. 
\end{abstract}

\section{Introduction}

The quantum optical processes and quantization of the electromagnetic (EM) field
in lossy structures exhibiting interference
have been widely studied during the last few decades especially in layered structures
\cite{Knoll1991,Allen1992,Huttner1992,Barnett1995,Matloob1995,Matloob1996}.
Describing the spatial evolution even in the most relevant quantum approaches
based on the input-output relation formalism (IORF) of the photon creation
and annihilation operators and direct quantization of Maxwell's equations
has been incomplete in the sense that they do not provide a unique way to determine
photon number. This is reflected by the observation that the vector potential and
electric field operators obtained by using the IORF obey the well-known canonical
commutation relation for an arbitrary choice of normal
mode functions as expected \cite{Barnett1995,Matloob1995}, but the commutation
relations of the ladder operators do not. The anomalous commutation relations
of the ladder operators were studied in several reports
\cite{Raymer2013b,Barnett1996,Ueda1994,Aiello2000,Stefano2000} but no clear resolution for the
anomalies was found apart from reaching a consensus that the anomalies were
irrelevant as long as the field commutation relations and classical field
quantities were well defined. Since then, the IORF has mainly been applied
in calculating the classical field quantities until the very recent suggestions that,
despite the early interpretations, the ladder operators and their commutation
relations might in fact relate to measurable physical properties
\cite{Collette2013} and finding well derived operators can significantly
simplify the description of energy transfer \cite{Partanen2014a}.

We have recently introduced photon ladder operators associated with the vector potential
in a way that is consistent with the canonical commutation relations
and gives further insight on the local effective photon number, thermal balance, and the
formation of the local thermal equilibrium \cite{Partanen2014a}. 
One interesting result is that the field at each point in a vacuum cavity couples differently to 
the surrounding material layers suggesting a position-dependent local photon number 
and temperature. The position dependence is expected to be directly observable using
experimental setups in which the field-matter interaction is dominated by the coupling
to the electric field.

In this paper, we apply our quantum fluctuational electrodynamics approach to
describe the EM field in lossless and lossy cavity geometries.
We start by a short review of the theoretical background to the EM field
quantization, photon operators, and their relation to thermal balance. This is followed by 
applying the presented theoretical concepts to study the position and photon energy
dependence of the effective field temperature, electrical contribution of the local density of EM states,
Poynting vector, and net emission rate in two cavity geometries: a vacuum
cavity formed between two semi-infinite thermal reservoir media at different temperatures,
and a similar structure where the cavity medium is lossy.

\section{Field quantization}
\label{sec:theory}

For completeness of the theoretical foundations of the present work, we first give an overview
of the results originally presented by Matloob \emph{et al.} \cite{Matloob1995}
to lay the ground for our recently defined ladder and photon-number
operators \cite{Partanen2014a}. After that the ladder and photon-number
operators are presented and their relation to the Poynting vector and
thermal balance is discussed.

\subsection{Overview of the noise operator formalism for EM field quantization}

The electromagnetic waves are considered to propagate parallel to the $x$ axis with their
transverse electric and magnetic field operators $\hat E(x,t)$ and $\hat B(x,t)$
parallel to the $y$ and $z$ axes, respectively. The field operators are related
to the vector potential operator $\hat A(x,t)$ by the relations \cite{Matloob1995}
\begin{align}
 \hat E^+(x,\omega) & = i\omega\hat A^+(x,\omega)\label{eq:EfromA},\\
 \hat B^+(x,\omega) & = \frac{\partial}{\partial x}\hat A^+(x,\omega)\label{eq:BfromA}
\end{align}
for positive frequencies. The negative frequency parts $\hat E^-(x,\omega)$,
$\hat B^-(x,\omega)$, and $\hat A^-(x,\omega)$ are Hermitian conjugates
of the positive frequency parts.

The field operators satisfy the frequency domain Maxwell's equations and,
when the relations in Eqs.~\eqref{eq:EfromA} and \eqref{eq:BfromA} are used,
the vector potential operator can be shown to satisfy the
one-dimensional nonhomogeneous Helmholtz equation \cite{Matloob1995}
\begin{equation}
\frac{\partial^2}{\partial x^2}\hat A^+(x,\omega)+\frac{\omega^2n(x,\omega)^2}{c^2}\hat A^+(x,\omega)=-\mu_0\hat J_\mathrm{em}(x,\omega),
\label{eq:Helmholtz}
\end{equation}
where $\mu_0$ is the permeability of vacuum,
$n(x,\omega)$ is the refractive index of the medium, and
$\hat J_\mathrm{em}(x,\omega)=j_0(x,\omega)\hat f(x,\omega)$ is a Langevin noise current operator
presented in terms of the scaling factor
$j_0(x,\omega)$ and the modified Langevin force operator $\hat f(x,\omega)$ which is a
bosonic field operator defined through the following commutation relations
$[\hat f(x,\omega),\hat f^\dag(x',\omega')] = \delta(x-x')\delta(\omega-\omega')$
and $[\hat f(x,\omega),\hat f(x',\omega')] =[\hat f^\dag(x,\omega),\hat f^\dag(x',\omega')]=0$.
The scaling factor of the Langevin noise current operator is given by $j_0(x,\omega)=\sqrt{4\pi\hbar\omega^2\varepsilon_0\mathrm{Im}[n(x,\omega)^2]/S}$,
where $\hbar$ is the reduced Planck's constant, $\varepsilon_0$ is the permittivity
of vacuum, and $S$ is the area of quantization in the $y$-$z$ plane \cite{Matloob1995}.
The quantization area only affects the scaling of the field quantities and in the calculations we set $S=1$ m$^2$.
The magnitude of the scaling factor has been determined by requiring that the vector potential
and electric field operators obey the canonical equal-time commutation relation
as detailed in Ref.~\cite{Matloob1995}\hspace{0.1cm}.

The solution to Eq.~\eqref{eq:Helmholtz} can be written in terms of the Green's function
of the Helmholtz equation as
\begin{equation}
 \hat{A}^+(x,\omega) = \mu_0\int_{-\infty}^\infty j_0(x',\omega) G(x,\omega,x')\hat f(x',\omega)dx'.
 \label{eq:solutionA}
\end{equation}
The Green's function depends on the problem geometry via the refractive index of the medium.
The Green's function for the studied two interface geometry is given in Ref.~\cite{Partanen2014a}\hspace{0.1cm}.
In order to write the field operators in compact forms, we define the scaled forms
of the Green's functions:
\begin{align}
 G_\mathrm{A}(x,\omega,x') = & \mu_0j_0(x',\omega)G(x,\omega,x')\label{eq:GA},\\[10pt]
 G_\mathrm{E}(x,\omega,x') = & i\mu_0\omega j_0(x',\omega)G(x,\omega,x')\label{eq:GE},\\[8pt]
 G_\mathrm{B}(x,\omega,x') = & \frac{i\mu_0\omega n(x,\omega)}{c}j_0(x',\omega)[G_\mathrm{R}(x,\omega,x')-G_\mathrm{L}(x,\omega,x')]\label{eq:GB},
\end{align}
where $G_\mathrm{E}(x,\omega,x')$ and $G_\mathrm{B}(x,\omega,x')$ are obtained from $G_\mathrm{A}(x,\omega,x')$
by using Eqs.~\eqref{eq:EfromA} and \eqref{eq:BfromA}, and 
$G_\mathrm{R}(x,\omega,x')$ and $G_\mathrm{L}(x,\omega,x')$ are the right and left
propagating parts of the Green's function identified from the factors
$e^{ikx}$ and $e^{-ikx}$,
respectively. In this paper, when treating lossless semi-infinite media,
an infinitesimal imaginary part of the refractive index is assumed in $j_0(x,\omega)$ and $G(x,\omega,x')$
and it is set to zero after calculating the integrals. Using the above definitions the
vector potential operator is given by $\hat{A}^+(x,\omega) = \int_{-\infty}^\infty G_\mathrm{A}(x,\omega,x')\hat f(x',\omega)dx'$
and corresponding expressions are valid for $\hat{E}^+(x,\omega)$ and $\hat{B}^+(x,\omega)$.
In time domain the fields are given by the inverse Fourier transforms of the frequency domain operators
as $\hat{A}(x,t)= \frac{1}{2\pi}\int_0^\infty\hat A^+(x,\omega)e^{-i\omega t}d\omega + \mathrm{H.c.}$,
where the Hermitian conjugate is the negative frequency part. Again, $\hat E(x,t)$ and $\hat B(x,t)$
are obtained from similar expressions.

The electric displacement field operator $\hat{D}(x,t)$ and the magnetic field strength operator $\hat{H}(x,t)$,
needed, e.g., in calculating the Poynting vector, are obtained from 
the electric field and magnetic field density operators using the constitutive relations
$\hat{D}^+(x,\omega)=\varepsilon_0\varepsilon(x,\omega)\hat{E}^+(x,\omega)$
and $\hat{B}^+(x,\omega)=\mu_0\mu(x,\omega)\hat{H}^+(x,\omega)$,
where $\varepsilon(x,\omega)$ and $\mu(x,\omega)$ are the 
position-dependent relative permittivity and permeability of the medium  \cite{Novotny2006}.

\subsection{Ladder and photon-number operators}
\label{sec:operators}
In any quantum electrodynamics (QED) description, the canonical commutation relations are satisfied for
field quantities, i.e., $[\hat A(x,t),\hat E(x',t)]=-i\hbar/(\varepsilon_0S)\delta(x-x')$ \cite{Scheel1998},
but the same is not generally true for the canonical commutation relations of the ladder operators.
The dominant approach in evaluating the ladder operators has been to separate the
field operators obtained from QED either into the left and right 
propagating normal modes or into the normal modes related to the left and right inputs
and the corresponding ladder operators so that the vector potential can be written as
$\hat A^+(x,\omega)=u_\mathrm{R}(x)\hat a_\mathrm{R}(\omega) + u_\mathrm{L}(x)\hat a_\mathrm{L}(\omega)$
\cite{Barnett1996,Gruner1996,Aiello2000}.
This is tempting in view of the analogy with classical EM, but in most cases results in
ladder operators that are not unambiguously determined due to the possibility to scale
the normal modes nearly arbitrarily. Also divisions accounting for
the noise contribution and more physically transparent interpretations \cite{Stefano2000}
have been reported, but they do not give the canonical commutation relations for the ladder
operators either.

We have adopted a different starting point to preserve the
canonical commutation relation $[\hat a(x,\omega),\hat a^\dag(x,\omega')]=\delta(\omega-\omega')$
by defining the photon annihilation operator to be proportional to the total vector potential
operator and normalizing it so that the commutation relation is fulfilled \cite{Partanen2014a}.
Our approach gives the photon annihilation operator as
\begin{equation}
 \hat a(x,\omega)=\sqrt{\frac{\varepsilon_0\omega}{2\pi^2\hbar\rho(x,\omega)}}\hat A^+(x,\omega),
 \label{eq:totalfielda}
\end{equation}
where we have used the conventional definition for the electrical contribution of the
local density of EM states (electric LDOS) as \cite{Joulain2003}
\begin{equation}
 \rho(x,\omega)=\frac{2\omega}{\pi c^2S}\mathrm{Im}[G(x,\omega,x)].
 \label{eq:ldos}
\end{equation}

The photon-number operator is given in
terms of the ladder operators as
$\hat n(x,\omega)=\int\hat a^\dag(x,\omega)\hat a(x,\omega')d\omega'$
and its expectation value is expressed in terms of the Green's function as
\begin{equation}
 \langle\hat n(x,\omega)\rangle=\frac{\varepsilon_0\omega}{2\pi^2\hbar\rho(x,\omega)}\int_{-\infty}^\infty |G_\mathrm{A}(x,\omega,x')|^2\langle\hat\eta(x',\omega)\rangle dx'.
 \label{eq:photonnumber}
\end{equation}
Here we have defined a source field photon-number operator as
$\hat \eta(x,\omega) = \int\hat f^\dag(x,\omega)\hat f(x',\omega')\,dx'd\omega'$
and assumed that the noise operators at different positions and at different frequencies
are uncorrelated so that the source field photon-number expectation value at
position $x$ of a thermally excited medium is
\begin{equation}
 \langle\hat \eta(x,\omega)\rangle = \frac{1}{e^{\hbar\omega/(k_\mathrm{B}T(x))}-1},
 \label{eq:sourcefieldn}
\end{equation}
where $k_\mathrm{B}$ is the Boltzmann constant and
$T(x)$ is the position-dependent temperature of the medium.
In the case of thermal fields the photon-number operator in Eq.~\eqref{eq:photonnumber}
also allows one to calculate an effective local field temperature for the electric field as
\begin{equation}
 T(x,\omega)=\frac{\hbar\omega}{k_\mathrm{B}\ln[1+1/\langle\hat n(x,\omega)\rangle]}.
 \label{eq:temperature}
\end{equation}

In the above definition, the ladder
operators are essentially defined to be proportional to the total
vector potential (and therefore also the electric field) operator
including the contribution from all the source points. Furthermore,
the photon-number operator can be considered as an effective
operator that partly reflects the physical properties of the
electric field. For instance, under certain nonequilibrium
conditions studied in Sec.~\ref{sec:results}, the photon number
can oscillate due to the interference seen in the electric
fields. It is also found that without the above exact form
for the position-dependent normalization coefficient
the resulting photon-number expectation value
oscillates near material interfaces at thermal equilibrium. The properly
normalized annihilation operator defined in Eq.~\eqref{eq:totalfielda}
always results in a photon-number
expectation value that is constant everywhere
at thermal equilibrium.

\subsection{Poynting vector and thermal balance}
\label{sec:balance}

For additional physical insight, we will compare the position dependence of the photon-number
expectation value following from Eq.~\eqref{eq:photonnumber}
to the well-known energy flux given by the Poynting vector.
The expectation value of the one-dimensional quantum optical Poynting vector operator
is given in terms of the electric and magnetic field Green's functions
and the source field photon-number expectation value as \cite{Partanen2014a}
\begin{equation}
 \langle\hat{S}(x,t)\rangle_\omega =\frac{1}{2\pi^2}\int\mathrm{Re}[G_\mathrm{E}^*(x,\omega,x')G_\mathrm{H}(x,\omega,x')]\langle\hat\eta(x',\omega)\rangle dx',
 \label{eq:poynting}
\end{equation}
where $G_\mathrm{H}(x,\omega,x')=G_\mathrm{B}(x,\omega,x')/(\mu_0\mu(x,\omega))$.
The brackets denote the expectation value over all states
resulting in the source field photon-number expectation value of Eq.~\eqref{eq:sourcefieldn},
and the subscript $\omega$ denotes the spectral component of the Poynting vector,
i.e., the integrand when the total Poynting vector is expressed as an integral
over positive frequencies.

For a better understanding of the introduced photon-number concept it is essential to consider
its contribution to local energy balance at position $x$. The Poynting theorem
relates the local power dissipation and generation
to the current density and electric field at position $x$ and to
the divergence of the Poynting vector \cite{Jackson1999}. Therefore, in one dimension, the spectral
energy transfer rate $\langle Q(x,t)\rangle_\omega$, i.e., the spectral net emission between
the field and the medium at position $x$, is given by
\begin{equation}
 \langle Q(x,t)\rangle_\omega=\frac{\partial}{\partial x}\langle\hat S(x,t)\rangle_\omega=-\langle\hat J(x,t)\hat E(x,t)\rangle_\omega.
 \label{eq:conservation}
\end{equation}
The current density term $\hat J(x,t)=\hat J_\mathrm{abs}(x,t)-\hat J_\mathrm{em}(x,t)$
consists of parts corresponding to absorption and emission.
The photon emission is described by the Langevin noise current operator
$\hat J_\mathrm{em}(x,\omega)=j_0(x,\omega)\hat f(x,\omega)$
introduced as a field source in Eq.~\eqref{eq:Helmholtz}.
The photon absorption term is of the form
$\hat J_\mathrm{abs}(x,\omega)=\varepsilon_0\omega\,\mathrm{Im}[n(x,\omega)^2]\hat E^+(x,\omega)$
so that the absorption rate is proportional to the square of the electric field,
or equivalently the electric field fluctuation.
Substituting the time domain current terms (i.e., the Fourier
transforms of the frequency domain terms) to Eq.~\eqref{eq:conservation} and calculating
the expectation values over source field photon-number states gives the
local net emission rate in terms of the photon numbers of
the source and the total electromagnetic fields as
\begin{equation}
 \langle Q(x,t)\rangle_\omega=\hbar\omega^2\mathrm{Im}[n(x,\omega)^2]\rho(x,\omega)[\langle\hat\eta(x,\omega)\rangle-\langle\hat n(x,\omega)\rangle].
 \label{eq:divP}
\end{equation}
Equation \eqref{eq:divP} directly shows that the local net emission rate is
zero only if the material is lossless $(\mathrm{Im}[n(x,\omega)^2]=0)$, the electric LDOS is zero $[\rho(x,\omega)=0]$,
or the field is at local thermal equilibrium $[\langle\hat n(x,\omega)\rangle=\langle\hat\eta(x,\omega)\rangle]$.
Equation \eqref{eq:divP} also nicely separates the effect of temperature and
wave features in the local net emission rate: the effect of temperature is described
by the photon-number operators and the effect of wave features is described by the
imaginary part of the Green's function.
In resonant systems where the energy exchange is dominated by a narrow frequency band,
condition $\langle\hat Q(x,\omega)\rangle_\omega=0$ can be used
to  approximately determine the steady state temperature of a weakly interacting resonant
particle \cite{Bohren1998}. This leads to concluding that in order to reach a thermal
balance with the field, the particle must reach a temperature
that is equal to the effective field temperature so that the term
$\langle\hat\eta(x,\omega)\rangle-\langle\hat n(x,\omega)\rangle$ disappears.

\section{Results}
\label{sec:results}

To investigate the physical implications of the concepts presented in Sec.~\ref{sec:theory}
we study the electric LDOS, effective field temperature, Poynting vector, and net emission rate in two geometries:
a vacuum cavity formed between two semi-infinite thermal reservoir media at different temperatures,
and a similar structure where the cavity medium is lossy.

\subsection{Vacuum cavity}

\begin{figure}
\includegraphics[width=\textwidth]{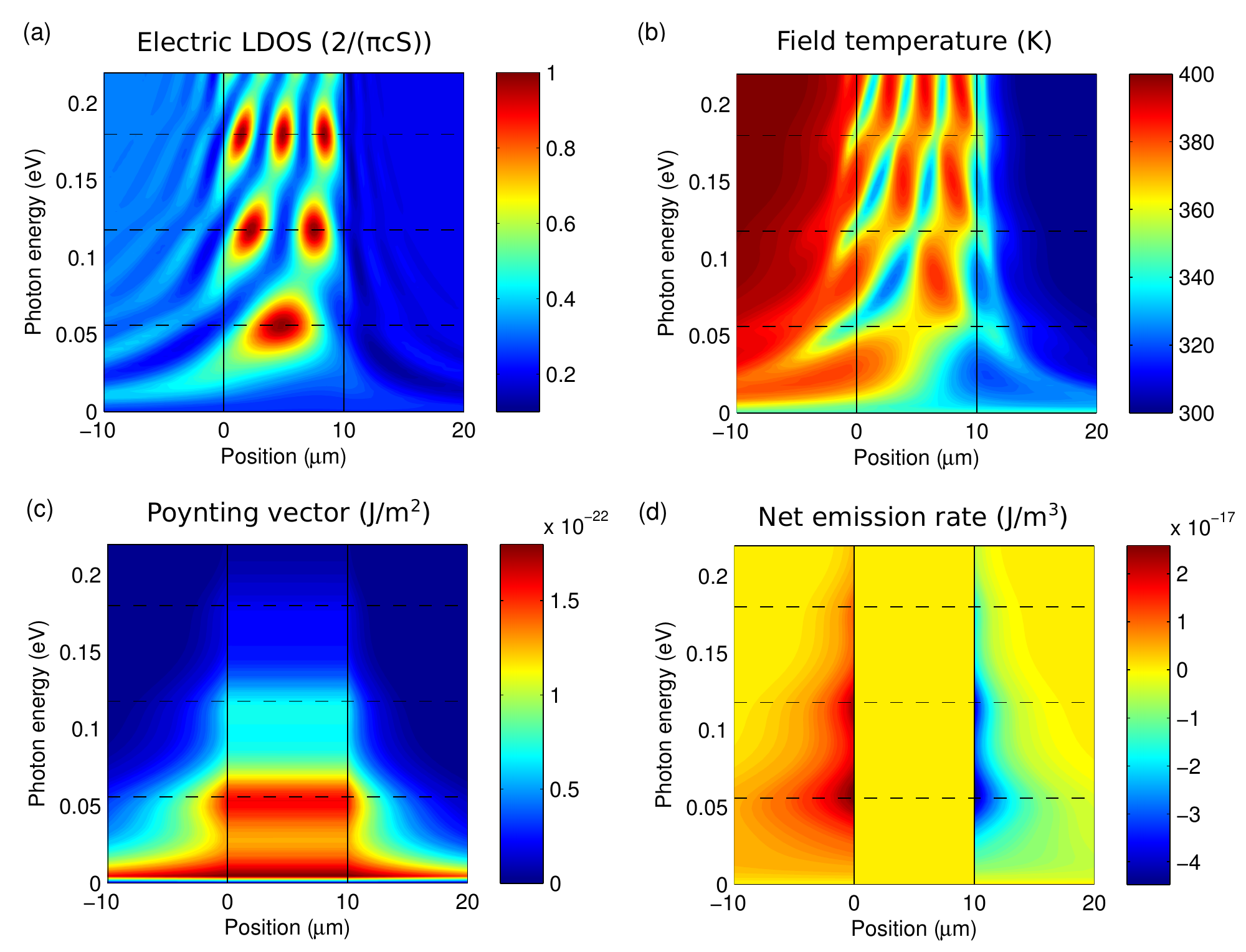}
\caption{\label{fig:lossless}(Color online) (a) Electric LDOS, (b) effective field temperature,
(c) Poynting vector, and (d) the net emission rate in the vicinity of a vacuum gap separating
lossy media with refractive indices $n_1=1.5+0.3i$ and $n_2=2.5+0.5i$ at temperatures 400 and 300 K.
Solid lines denote the boundaries of the cavity and dashed lines denote resonant energies.
The electric LDOS is given in the units of $2/(\pi c S)$.}
\end{figure}

The vacuum cavity structure consists of two semi-infinite media
with refractive indices $n_1=1.5+0.3i$ and $n_2=2.5+0.5i$ at temperatures
$T_1=400$ K and $T_2=300$ K, separated by a  $10$ $\mu$m vacuum gap.
Figure \ref{fig:lossless} shows the electric LDOS,
effective field temperature, Poynting vector, and net emission rate as a function of position
and photon energy for the vacuum cavity geometry.
The electric LDOS plotted in Fig.~\ref{fig:lossless}(a) oscillates in the vacuum and saturates to
constant values in the lossy media far from the interfaces. The oscillation
of the electric LDOS inside the cavity is strongest at resonant energies $\hbar\omega=0.056$ eV ($\lambda=22.1$ $\mu$m),
$\hbar\omega=0.118$ eV ($\lambda=10.5$ $\mu$m), and $\hbar\omega=0.180$ eV ($\lambda=6.89$ $\mu$m).
The electric LDOS naturally has one, two, and three peaks at the energies of the first, second,
and third resonances. The oscillations in the electric LDOS directly reflect
the Purcell effect and position-dependent emission rate of particles placed in the cavity.

The effective field temperature defined using Eq.~\eqref{eq:temperature} is plotted in
Fig.~\ref{fig:lossless}(b). It has a strong position dependence
and it oscillates both in the vacuum and inside the lossy media. The position dependence originates
from the unequal coupling to the two thermal reservoirs. In the lossy media
the oscillations are damped and the effective field temperature saturates to constant
values far from the interfaces. The distance for the damping depends on the
photon energy and the material absorptivity. In the case of the second resonant energy, the damping takes place
over a distance of 10 $\mu$m. For smaller photon energies, the distance for the
damping is longer and for larger photon energies the damping takes place over a
shorter distance. The oscillations of the effective field temperature in the vacuum
are expected to be related to defining the ladder operators so 
that they are proportional to the vector potential and the electric field. This implies 
that the resulting effective photon number and field temperature are in fact quantities
that mainly reflect the features related to electric field and field-matter interactions
involving electrical dipoles.

The Poynting vector is plotted in Fig.~\ref{fig:lossless}(c) and its derivative, which equals
the net emission rate also obtained through Eq.~\eqref{eq:divP},
is plotted in Fig.~\ref{fig:lossless}(d). In the vacuum gap,
the Poynting vector is constant and the net emission rate is zero with respect to the
position since there is no interaction.
The positivity of the Poynting vector denotes
net energy transfer towards the medium at lower temperature.
Correspondingly, the positive (negative)
values of the net emission rate denote that the rate of photon emission (absorption)
outweighs the rate of photon absorption (emission).
Inside the lossy media, the Poynting vector and the net emission rate
asymptotically reach zero far from the interfaces.
The damping of the Poynting vector and the net emission rate
is faster in the right medium than in the left medium due
to the larger imaginary part of the refractive index, i.e.~the absorption coefficient.
The value of the Poynting vector and the net emission rate also go to zero at higher frequencies
since the source field photon-number expectation value is smaller compared
to the photon number of smaller frequencies as given by the Bose-Einstein distribution in Eq.~\eqref{eq:sourcefieldn}.
However, the value of the Poynting vector and the net emission rate slightly peak at resonant energies
since the total transmission coefficient of the cavity achieves its maximum values
at resonant energies, which increases the energy transfer across the cavity.

\subsection{Lossy cavity}

As a second example we replace the vacuum cavity with lossy material having a refractive index
$n_\mathrm{c}=1.1+0.1i$ and calculate the electric LDOS, effective field temperature, Poynting
vector, and the net emission rate in the cavity. As in the case of the lossless cavity structure,
the source field temperatures of the two semi-infinite reservoirs are $T_1=400$ K and $T_2=300$ K.
In contrast to the lossless cavity, the lossy medium inside the cavity acts as and additional
field source emitting photons. In this work, we calculate the temperature of the lossy medium
inside the cavity self-consistently so that the photon emission equals absorption
at every point. This also means that other heat conduction mechanisms than radiation are neglected.

\begin{figure}
\includegraphics[width=\textwidth]{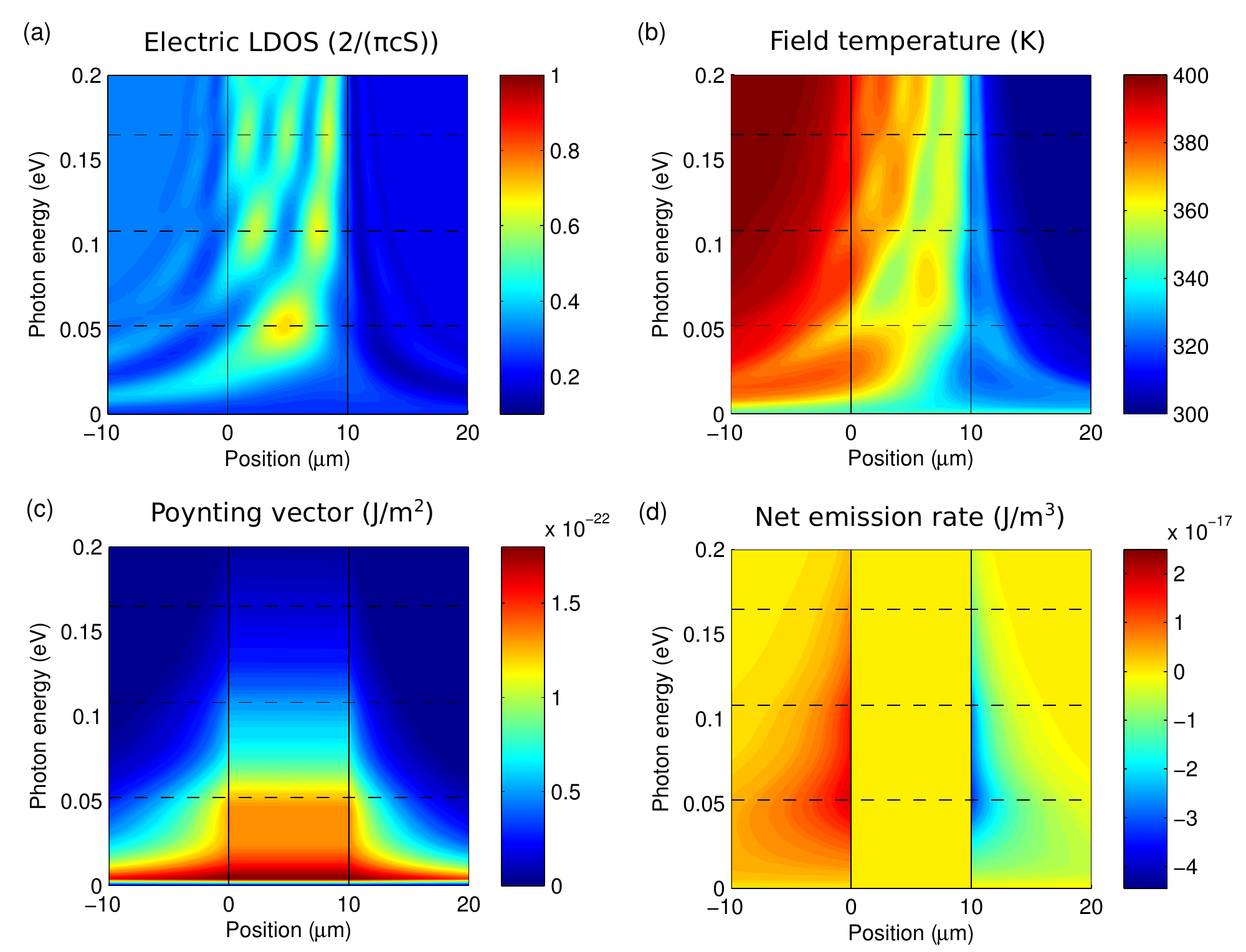}
\caption{\label{fig:lossy}(Color online) (a) Electric LDOS, (b) effective field temperature,
(c) Poynting vector, and (d) the net emission rate in the vicinity of a lossy material layer
with refractive index $n_c=1.1+0.1i$ separating lossy media with refractive indices
$n_1=1.5+0.3i$ and $n_2=2.5+0.5i$ at temperatures 400 and 300 K.
Solid lines denote material interfaces and dashed lines denote resonant energies.
The electric LDOS is given in the units of $2/(\pi c S)$.}
\end{figure}

The electric LDOS for the lossy cavity structure is plotted in Fig.~\ref{fig:lossy}(a).
The electric LDOS again oscillates inside the cavity, but the peak values are smaller due to losses. 
In the lossy media on the left and right, the oscillation saturates to
constant values far from the interfaces. The oscillation
of the electric LDOS inside the cavity is strongest at resonant energies $\hbar\omega=0.052$ eV ($\lambda=23.8$ $\mu$m),
$\hbar\omega=0.108$ eV ($\lambda=11.5$ $\mu$m), and $\hbar\omega=0.165$ eV ($\lambda=7.53$ $\mu$m).
As in the case of the lossless cavity in Fig.~\ref{fig:lossless}(a),
the number of peaks in the electric LDOS at different resonant energies
corresponds to the ordinal number of the resonance.

The effective field temperature is shown in Fig.~\ref{fig:lossy}(b). The effective field temperature
has a less pronounced position dependence inside the cavity when compared to the case of the lossless
cavity in Fig.~\ref{fig:lossless}(b) since the oscillations are now damped due to the losses.
In the lossy media on the left and right, the oscillations are again damped
and the effective field temperature saturates to a constant
value far from the interfaces. The distance for the damping depends on the
photon energy corresponding to the damping distance in the case of
the vacuum cavity geometry since the left and right lossy media are the same.

The Poynting vector plotted in Fig.~\ref{fig:lossy}(c) is constant with respect
to the position inside the cavity. This follows from calculating the source field temperature
of the medium inside the cavity self-consistently so that the
emission equals absorption. Since the emission equals absorption, the net emission
rate in Fig.~\ref{fig:lossy}(d) is zero inside the cavity as detailed in Eq.~\eqref{eq:divP}.
Therefore, the Poynting vector is necessarily constant inside the cavity.
Inside the surrounding lossy media, the Poynting vector
evolves as in the lossless cavity case so that
the Poynting vector and the net emission rate again
asymptotically reach zero far from the interfaces.
However, due to the losses inside the cavity the positions
of the resonant energies are not as easily seen
in the Poynting vector as in the
lossless cavity geometry in Fig.~\ref{fig:lossless}(c).
The absolute values of the net emission rates on the
left and right are lower when compared to the net emission
rate of the lossless cavity geometry in Fig.~\ref{fig:lossless}(d).
This is due to the losses inside the cavity which reduce
the energy transfer across the cavity.

The proposed position-dependent ladder and photon-number operators predict
that the effective field temperature oscillates with respect
to the positions as shown in Figs.~\ref{fig:lossless}(b) and \ref{fig:lossy}(b). 
In contrast to the field quantities, the effective photon number provides
a simple metric for finding the thermal balance formed due to interactions
taking place through the electric field and electric dipoles as detailed in
Eq.~\eqref{eq:divP}. Since the photon number
as defined in this work is also expected to be directly related both
to local temperature and rate of energy exchange taking place between the
electric field and the dipoles constituting the lossy materials, we 
expect that the predicted photon number and field temperature oscillations
can also be measured. A good candidate for such a measurement would be
a setup consisting of a high-quality-factor cavity
with a single photon detector.
Photon-number measurements have been demonstrated in such cavities \cite{Maitre1997},
but not in nonequilibrium conditions which is necessary for the oscillations of the effective
photon number and field temperature with respect to the position. If the cavity is asymmetric,
the photon-number oscillations are more easily observable
than in a symmetric cavity where the photon-number oscillations
can disappear at resonant frequencies.
As the taken approach is very general, generalizing the model
to three dimensions is expected to be straightforward.
In this paper we have investigated only thermal source fields in detail, but
the introduced operators are expected to enable also a much more general description
of the quantized fields obeying other kinds of quantum statistics.
For example, replacing the noise operators with operators describing partly
saturated emitters \cite{Hayrynen2010b} could result in a simple but realistic
quantum description of the generation of laser fields, and, furthermore,
the use of nonlinear source field operators \cite{Hayrynen2012b,Hayrynen2010c}
could even allow modeling single photon sources and detectors.

\section{Conclusions}
\label{sec:conclusions}
We have studied the application of the recently introduced position-dependent photon ladder operators
in lossless and lossy cavity structures. The introduced operators enable
a physically meaningful definition of an effective position-dependent
photon-number operator that has a very attractive and simple connection to the
electric field, the effective field temperature, and thermal balance of the system.
Our calculations show that the intracavity operators are superpositions of the fields originating from
different source points at the extracavity material leading to position-dependent effective photon
number and local temperature in nonequilibrium conditions.
The effective photon number and field temperature are
expected to be observable in measurements in which the field-matter
interaction is dominated by the coupling to the electric field. This essentially
differentiates our definition of the effective photon number from the conventional
position-independent definitions that do not take the physical properties of the
electric field similarly into account.
Possibly the greatest opportunities enabled by the
new formulation are in extending the description to quantum systems
that are not limited to thermal fields and, as a practical point of view,
using the approach into planning, e.g., efficient photon emitters and
energy conversion applications.

\begin{acknowledgments}
This work has in part been funded by the Academy of Finland and the Aalto Energy Efficiency Research Programme.
\end{acknowledgments}

\bibliographystyle{spiebib}
\bibliography{bibliography}

\end{document}